# Fast method of crosstalk characterization for HxRG detectors

**Elizabeth M. George[1*], Simon M. Tulloch[1], Derek J. Ives[1], and the ESO detector Group**

[1] European Southern Observatory, 85748 Garching, Germany

**Abstract.** HxRG detectors have crosstalk between amplifier channels at a scientifically relevant level. In principle, crosstalk signals can be fully calibrated and removed from data, but only if a full crosstalk matrix is measured for the detector. We present a fast method of crosstalk characterization that can be performed with most instrument calibration units; it requires only a flat-field illumination and window programming in the HxRG detectors. We show the crosstalk matrices obtained with this method for both fast and slow mode in an H2RG detector, and give examples of how this data can be used to tune detector operation parameters, feed back into the electronics design for the cryogenic pre-amplifiers, and be used in the data pipeline to remove crosstalk signal from scientific data.

**Keywords:** crosstalk, NIR detector, calibration, HxRG.

*****Elizabeth M. George,** E-mail: egeorge@eso.org

## 1   Introduction and goals

Hawaii x Reference Guide (HxRG) detectors, where the x stands for 1,2 or 4, denoting $1024^2$, $2048^2$, or $4096^2$ pixels, are widely used in astronomy. HxRG detectors are read out in blocks of vertical stripes on the detector. [1] These blocks are read out simultaneously by separate amplifiers, and a signal (such as a bright star) appearing in one of these blocks can couple to the other amplifiers during readout. This signal is known as amplifier crosstalk.

   Amplifier crosstalk is a familiar feature in H2RG devices, and it appears in scientific data in many on-sky instruments. As an example, the XSHOOTER instrument [2] is a cross-dispersed echelle spectrograph using a H2RG detector. In this instrument, the spectral lines appear as nearly-vertical tilted lines. Cross-talk from bright spectral lines then appear as "ghost-lines" in the other amplifier blocks, leading to erroneous line features in the spectra (see Figure 5 of [3] for an example of the crosstalk signal in an arc-lamp spectral image). The authors of [3] show that the crosstalk in each amplifier is a fraction (typically < 1%) of the time derivative of the original signal (e.g. the derivative in the order the pixels are read out), indicating that the coupling is likely capacitive.

   While in an ideal situation we would avoid introducing crosstalk in the first place, in reality there will always be some (possibly low) level of electronic crosstalk. However, even low levels of electronic crosstalk can affect scientific measurements. If the crosstalk coupling amplitude can be measured between all amplifier pairs, we can calibrate it and remove it from scientific data. We therefore developed a method of crosstalk char-acterization that can be performed with most instrument calibration units. The goals of this method are to:
1. Quickly characterize cross-talk between all amplifier pairs in HxRG devices.
2. Determine optimum detector operation parameters for minimized crosstalk.
3. Provide engineering feedback to electronics design.
4. Provide crosstalk correction for science data.

## 2 Parameterizing crosstalk

HxRG devices are read out in blocks. If the detector is N x N pixels, and there are $j$ amplifiers, then there are $j$ blocks with a number of pixels in each block totaling N/$j$ columns x N rows. For example, in a H2RG detector read out in 32 channel output mode, there are 32 blocks of 64 x 2048 pixels, while in a H4RG detector read out in 64 channel output mode, there are 64 blocks of 64 x 4096 pixels. The $j$ amplifiers are read out simultaneously, with the "fast" direction of read along the row. In each block, the 64 pixels of a single row are read in order before moving to the next row in the block. By default, the direction of read of the rows is in opposite directions for even and odd amplifiers. We will focus for the remainder of this document on the specific case of a H2RG detector read out in the default 32 output mode, though the method described here could be applied to other HxRG family devices and readout configurations.

We assume that the cross-talk signal is a fraction of the derivative of the perturbing signal, as in [3]. We use the index $j$ ($0 \leq j \leq 31$) for the amplifier in which we observe the crosstalk signal. The index $i$ ($0 \leq i \leq 63$) is the pixel index running over pixels in a single row of each block. Or more simply, if we assume that the crosstalk in every row of a single block behaves the same, then $i$ is the column within a single amplifier *in the order in which the columns are read*—note that in the default readout mode, this direction switches between odd and even amplifiers. Using this convention, the cross talk signal $X$ on amplifier $j$ with respect to a signal $S$ on the amplifier $k$ ($0 \leq k \leq 31$) is:

$$X_{k,j,i} = a_{k,j} (S_{k,i} - S_{k,i-1})$$

**Eqn. 1**

By inputting a signal $S_k$ that has a strong derivative between columns, and measuring the crosstalk signal $X_{k,j}$, it is possible to determine the crosstalk matrix $a_{k,j}$, which is a 32 x 32 scalar matrix.

## 3 Measurement Method

To measure $a_{k,j}$ requires a bright source to be placed completely on a single amplifier $k$, and the resulting crosstalk signal to be measured on all amplifiers for $j \neq k$. Ideally, the detector would be in the dark, and an entire detector column would be illuminated such that the cross talk signal could be measured in every row in identical conditions. How-ever, it is nearly impossible in any reasonable optical setup to illuminate single columns, or even groups of adjacent columns, while leaving the neighboring columns unilluminated—even in a specialized detector test facility. Furthermore, some detectors may need to be characterized in-situ inside of instruments, and not every instrument has the ability to place a source solely in a single amplifier of the detector (for example, X-SHOOTER, a cross-dispersed Echelle spectrograph, or the IFS mode of SPHERE, a Spectro-Polarimetric High-contrast Exoplanet REsearch instrument [4]).

We therefore employ an alternative method to measure $a_{k,j}$ which requires only a (roughly) flat-field illumination and programming of detector clock patterns to reset windows of pixels on the detector. Most instruments have flat-field illumination built into the instrument calibration unit, the window-reset programming is straightforward and easy to implement in the detector controller, and the edges of the windows are sharply defined to be along single columns. Finally, since the location of the reset window is programmed, the location of the crosstalk signal in the other amplifiers is known exactly, and the signal therefore does not need to be searched for and fit. We describe the method and test setup in this section.

### 3.1 Test setup

We placed an engineering grade H2RG device inside of a test cryostat at a temperature controlled to 70K with a stability of a few mK. This engineering grade device has several features. There are vertical stripes along the columns on the right side of the detector affecting amplifiers 0-2. There are a few other small regions on the detector with horizontal or vertical striping that were excluded from the analysis. Additionally, amplifiers 23 and 24 are directly shorted together in slow mode and have a partial coupling in fast mode. Despite these features, the detector performance was adequate to validate this measurement technique.

Close to the detector is a version of the "ESO standard" cryogenic preamplifier board. The detector is controlled and read out with the standard ESO detector controller, NGC.

In this test cryostat, in front of the detector is a cold filter wheel with various positions: open, closed, bandpass and narrow-band filters from J-K bands. In order to get a pseudo-flat field, we performed all tests with a room temperature anodized aluminum plate in front of the cryostat window and a filter in the astronomical H-band. This resulted in approximately 30,000 counts in a 10s exposure, though the illumination was not perfectly uniform. Approximately half of each column was illuminated, allowing us to perform crosstalk measurements using the illuminated half of the detector. An ideal test setup would have a fairly uniform flat field illuminating the whole detector, but our test setup was adequate to validate this measurement technique.

### 3.2 Procedure

In HxRG devices, it is possible to arbitrarily define a pixel or rectangular window to reset, while leaving the other pixels in the array untouched. We created 32 different clock patterns, one for each amplifier, defining windows that are vertical stripes ~10 columns wide and spanning all rows of the detector. When this "large window" is reset, it creates a sharp change in signal between a detector column and its neighbor. We use this as an "on" image. To create an "off" image, we use 32 additional clock patterns in which only the first row of the same 10 columns (a "small window") is reset, leaving the remaining rows untouched.. To obtain the data used for the remainder of this paper, our procedure was as follows:

1) Set flat field source to a level that reaches ~60,000 electrons in 10 seconds. This value was chosen to be large enough that the crosstalk signal would be obvious, but still in the linear region of the detector.
2) For amplifiers k = 0 to 31 create an image with a window reset:
    a. Take "off" image (Reset array, Reset small window, Read, Integrate, Reset small window, Read) This looks like a flat field.
    b. Take "on" image (Reset array, Reset large window, Read, Integrate, Reset large window, Read) This looks like a flat field with a vertical stripe at ~zero counts
    c. Subtract the "off" image from the "on" image to create a processed image with ~0 counts everywhere except the reset window, where there will be ~-60,000 electrons signal. **Figure 1** shows a series of these images.

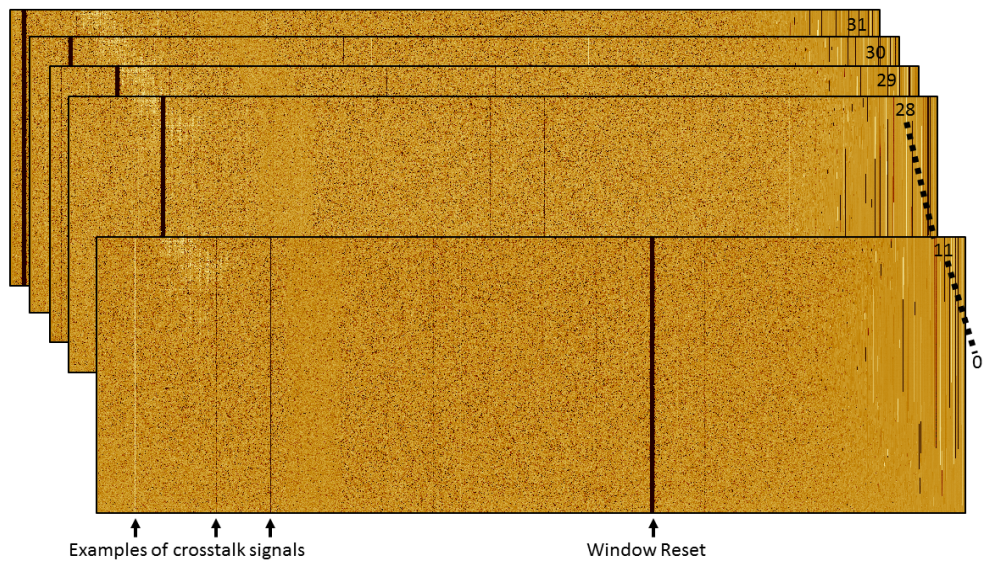

**Figure 1** Examples of the series of processed "On-Off" images produced in step 2. (Only 2048x652 pixels from the illuminated region are shown in each image). The numbers in the upper right corners indicate the amplifier with the reset window.

3) Average the light-sensitive pixels in each column to create a 1D plot of signal vs. column. In a clean device, this 1D map will have $\sqrt{2040}$ higher signal to noise than any single row of the 2D image. In our test setup with the engineering grade device only 652 rows could be used, and in some columns fewer. **Figure 2** shows an example of this 1D plot.

4) Record crosstalk in all amplifiers $j \neq k$ to populate $a_{k,j}$. (See next paragraph) **Figure 3** shows an example of these matrices for fast and slow mode measurements.
5) Repeat steps 1-4 for different read speeds and VBIASGATE values. (see section 4)

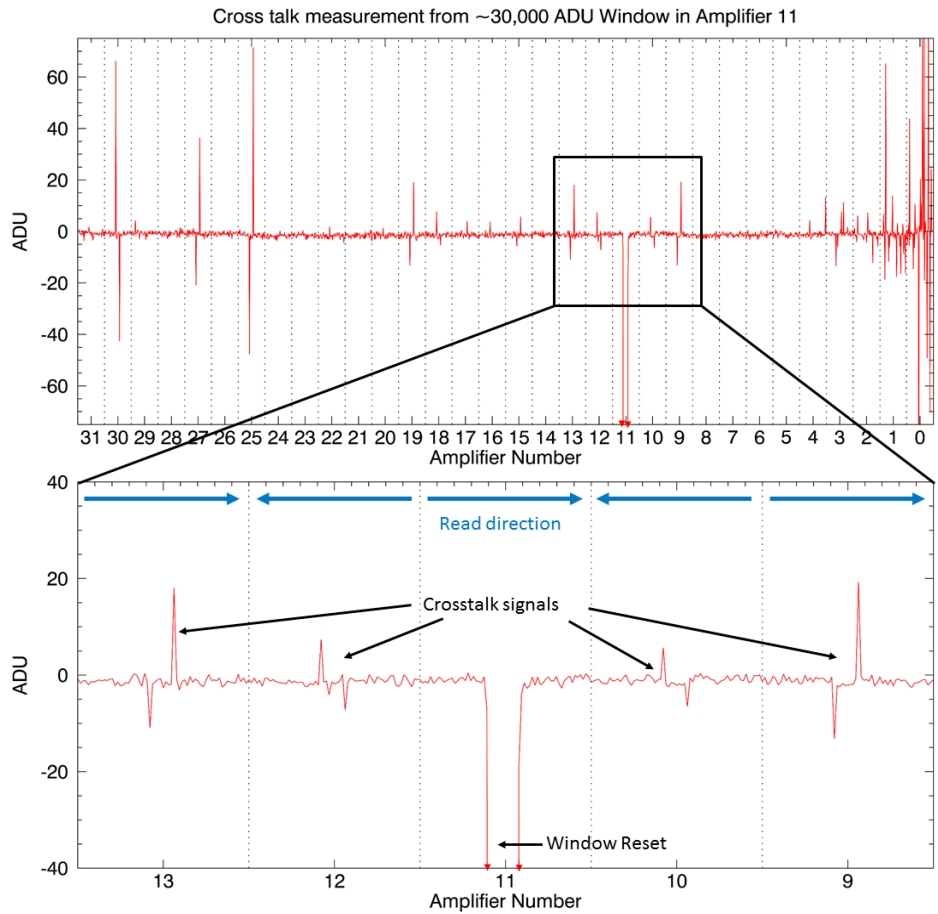

**Figure 2** Top: 1D plot produced in step 3. Bottom: Zoom of amplifier 11 and its nearest neighbors. The crosstalk signal is seen as spikes in single columns. The noise on the far right of the top plot is due to defects on this engineering grade device and is not due to crosstalk.

Note that in this procedure, it is relatively simple to record the crosstalk in step 4, as the change in signal is between two detector columns at the left edge of the window and two columns at the right edge of the window, and therefore the crosstalk signal appears in two single columns of each amplifier $j$ (one positive, and one negative). Each element of the crosstalk matrix is then approximately the amplitude of the spike above the zero-level divided by the depth of the reset window. We note that the positive and negative crosstalk spikes do not have identical amplitudes. This is likely due to the settling time constant during the window reset, as the first pixel within the window in each row is not reset as deeply as subsequent pixels within the window in each row.

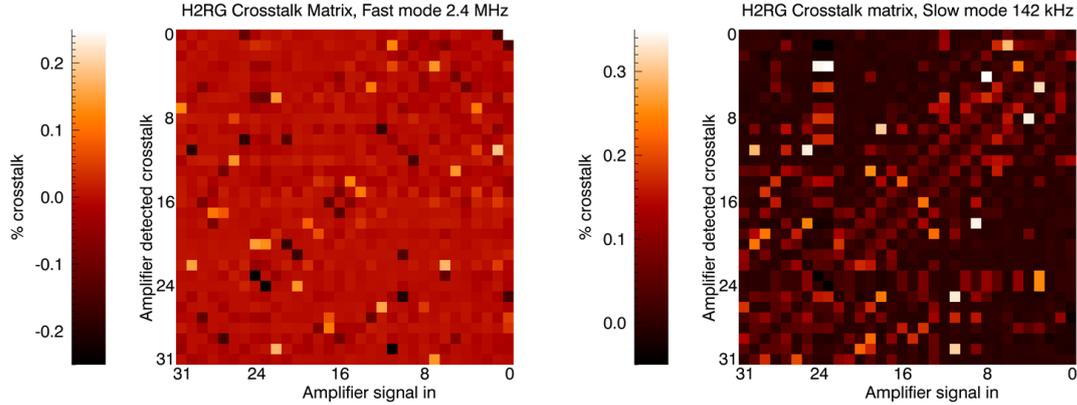

**Figure 3:** Crosstalk matrices produced in both fast and slow modes for an H2RG engineering device. The crosstalk signal seen in slow mode is all positive, while in fast mode, both positive and negative crosstalk was present. In this engineering grade device, amplifiers 23 and 24 are shorted together in slow mode, while in fast mode the coupling is incomplete.

We note here for completeness that this same procedure could be performed using only dark frames with different reset voltages for the whole detector and the reset window (e.g. by connecting VRESET to a clock instead of a DC voltage) instead of using flat field illumination, negating photon noise. However, many instrument controllers are not set up to easily allow the user to vary the reset voltage between resets, making the flat field illumination method the easiest choice for currently operating instruments.

## 4  Crosstalk matrix results

We performed several measurements using the procedure in section 3.2. We tested speeds ranging from 83-142 kHz in slow mode (slow unbuffered outputs) and 1-2.4 MHz in fast mode (fast buffered outputs). Additionally, we tested the effect of changing the current through the source follower, by changing VBIASGATE between values of 2.1-2.3 V. Examples of the crosstalk matrices obtained are shown in **Figure 3**. The fast mode crosstalk matrix shown is measured at a speed of 2.4 MHz with a VBIASGATE voltage of 2.3V, while the slow mode matrix shown is measured at a speed of 142 kHz and a VBIASGATE voltage of 2.2V. The crosstalk behavior with read speed and VBIASGATE voltage is shown in **Figure 4** for a single coefficient of the crosstalk matrix, however, the behavior is representative.

The properties of the crosstalk matrices will be system (preamplifiers, readout electronics, cables) and detector dependent. For the system and detector we tested the properties of the matrices. Results are listed below:
1. The crosstalk matrix is (mostly) symmetric.
2. There are many strong crosstalk signals far off-diagonal, though they are not always the same in fast and slow mode.

3. Strong crosstalk signals are positive in slow mode, and both positive and negative in fast mode.
4. On diagonal, there is more "pair" coupling in fast mode (e.g. pairs like 14/15 couple equally to each other), and "nearest neighbors" coupling in slow mode (e.g. a signal in one amplifier like 15 couples equally to its nearest neighbors, 14 and 16).
5. Crosstalk increases with increasing speed and decreases with increasing current through the source-follower. This confirms the Finger et al. 2008 result. (Note that increasing VBIASGATE voltage means decreasing current.)

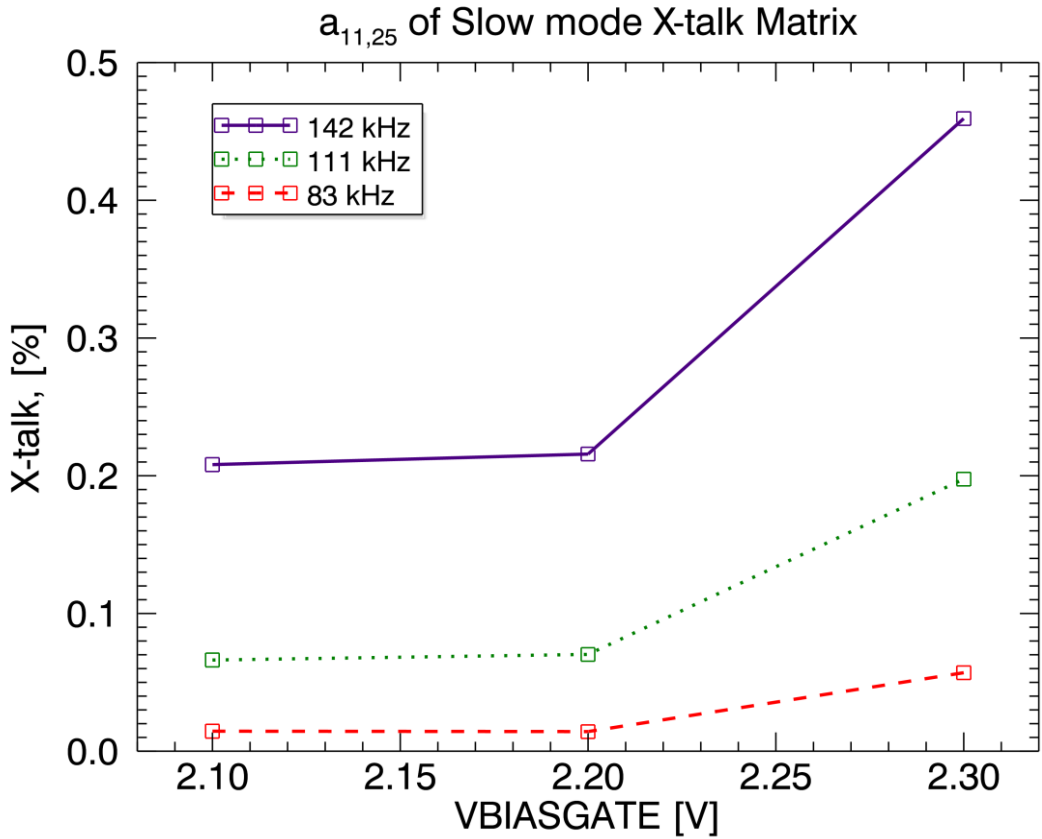

**Figure 4:** X-talk matrix element $a_{11,25}$ for slow mode as a function of VBIASGATE voltage. The solid line indicates data taken at a pixel speed of 142 kHz, the dotted line at 111 kHz, and the dashed line at 83 kHz.

We note that in HxRG detectors it is possible to read out in slow mode using the slow unbuffered outputs or the slow buffered outputs. The crosstalk is expected to be reduced further by using the buffered outputs, which we have confirmed with initial lab tests. This can be taken into consideration for future preamp design, though of course it cannot be easily changed in existing systems.

## 5 Engineering feedback

The measured crosstalk matrices can be used as engineering feedback to identify problems in the design. The symmetry of the matrix means that a signal in amplifier $k$ would cause crosstalk in amplifier $j$ with the same intensity as the crosstalk in amplifier $k$ caused by the same signal in amplifier $j$. There were a few exceptions to this symmetry, and these were traced to tracks on the preamplifier board where a pre-amplified signal ran parallel to an unamplified signal for considerable length, resulting in asymmetric coupling. In fact, many of the strong off-diagonal components of the crosstalk matrix could be directly traced to the pre-amplifier board design (see **Figure 5**). The new re-designed boards will have well separated tracks and shielding against capacitive coupling to prevent this type of crosstalk.

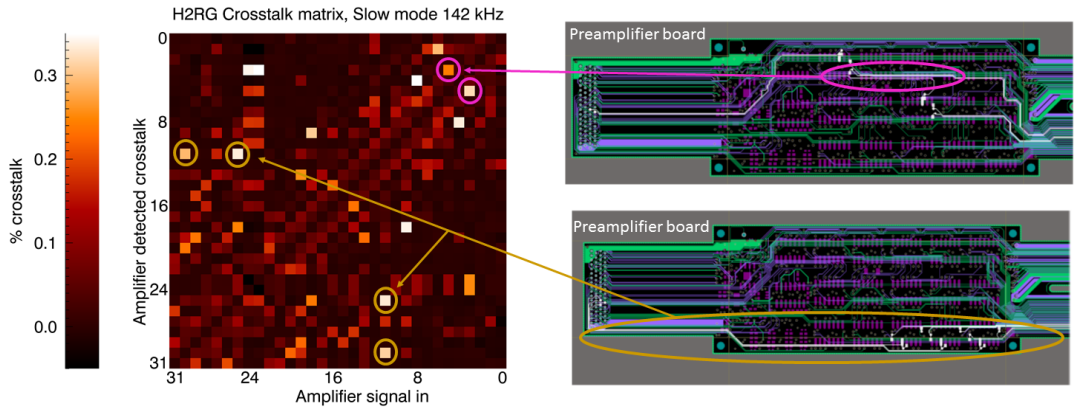

**Figure 5:** Examples of crosstalk signal traced to the preamplifier board. *Pink circles (top board):* Asymmetric coupling of amplifier 4 to amplifier 6 is due to the single-ended track of amplifier 6 coming from the detector running adjacent to the preamplified differential pair tracks of amplifier 4. *Yellow circles (bottom board):* Symmetric coupling of amplifiers 25 and 30 to amplifier 11 is due to the single-ended tracks from amplifiers 25 and 30 running adjacent on either side of the single-ended track from amplifier 11.

## 6 Crosstalk in science data

Crosstalk has been reported in several ESO instruments using H2RG devices. For example, KMOS is an integral field spectrograph where spectra are dispersed along the detector columns. A bright star dispersed along one amplifier can result in several crosstalk signals in other parts of the detector. When the field of view is reconstructed, these crosstalk signals look like erroneous positive and negative sources (see **Figure 6**).

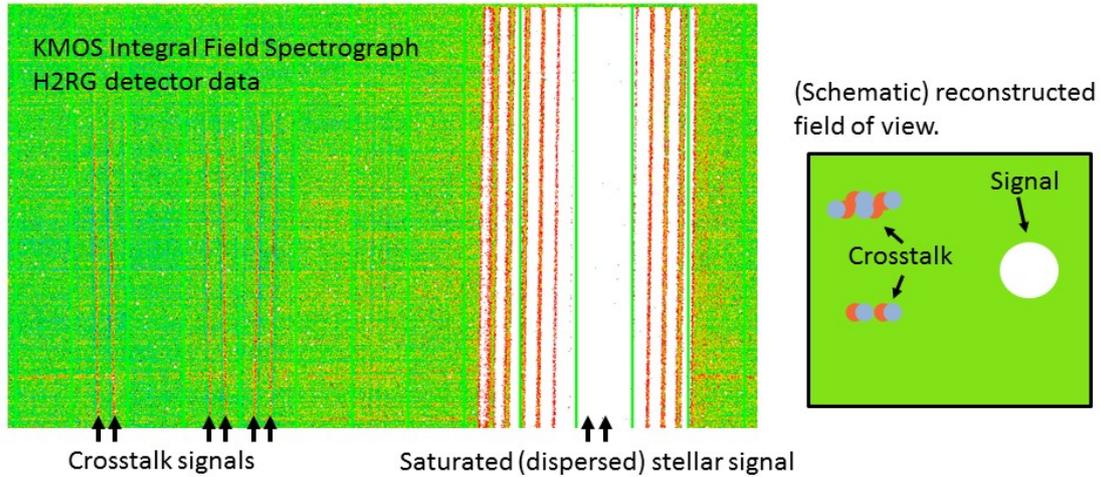

**Figure 6**: *Left:* An example of crosstalk signal seen in the KMOS integral field spectrograph. Stars are dispersed along the columns of the detectors, and the crosstalk signals appear as the derivatives of this star light in other amplifiers. The amplifier boundaries are delineated with the green lines. *Right*: A schematic of how a reconstructed IFS FOV with this crosstalk signal might appear. Note that the crosstalk signals appear as both positive and negative sources, and these crosstalk signals can appear in unexpected places in the FOV depending on how the field slices are arranged on the detector.

For future instruments such as HARMONI (also an IFS like KMOS), a crosstalk matrix can be produced using this method for the final instrument operation configuration (including all cables, electronics, and operation parameters). This crosstalk matrix can be optionally applied in the data analysis pipeline for observations in which there is expected to be a large difference in signal between the brightest and faintest objects in a field. For crosstalk at the 0.1% level, the crosstalk resulting from any objects in a field would be 7.5 magnitudes fainter. Crosstalk would be most noticeable when produced by the brightest objects in the field. In an IFS or multi-object spectrograph, there can be large differences in magnitude between objects in the same observation.

Crosstalk is an even bigger issue for instruments designed to do high contrast observations, such as VLT SPHERE. The SPHERE team decreased the effects of electrical crosstalk by changing detector operation parameters, including increasing the current through the source follower by lowering VBIASGATE from 2.4V to 2.2V and halving the readout speed to below 100 kHz. In addition, an electrical cross-talk correction routine was developed for SPHERE which further reduces the crosstalk to negligible levels [4]. While this effectively increased the achievable contrast ratio with the old detector parameters, the correction routine is no longer regularly used in the science data reduction as the changes in operation parameters reduced the crosstalk by a factor of 35 to a level where only 1-2 counts of crosstalk signal are expected for a full well exposure. [5]

## 7   Conclusions and future work

We have developed a fast method of crosstalk characterization that can be used with most instrument calibration units for in-situ crosstalk characterization. The resulting crosstalk matrices can be used for engineering feedback into the electronics design, de-tector operation parameter optimization, and correction of crosstalk in scientific data. We plan to repeat this measurement on new science grade devices for future VLT and ELT instruments using new preamplifier designs, at a range of speeds and bias voltage levels. These measurements will be used to optimize the detector performance for low crosstalk. Our tests show that lower crosstalk can be achieved by decreasing VBIASGATE and readout speed. In slow mode lower crosstalk is also expected by using the slow buffered outputs.

The crosstalk matrices produced with this method can be provided to the instrument analysis pipelines for use in correcting crosstalk in science data. While instruments such as SPIRou use an average crosstalk signal correction [6], our work shows that the coupling is amplifier-pair specific, and therefore for optimal correction the full matrix must be used. SPHERE has developed and implemented a crosstalk correction algorithm using a partial matrix based on measurements made using optical illumination. [5] The correction was implemented into the DRH recipes [4] which improved the science data (particularly for the old detector settings), [5] demonstrating that the correction can be a valuable tool for scientific instruments.

## Acknowledgments

Thanks to Ryan Houghton and HARMONI consortium for providing the example of crosstalk signal in KMOS data shown in **Figure 6**. We would also like to acknowledge the SPHERE team for their technical note showing improvements in the contrast ratio by implementing a partial correction using **Eqn. 1**. Finally, many thanks to the rest of the ESO detector group for supporting this work in many ways, in particular the continuous development of the ESO NGC system allowing us full and flexible control of detector operation.

**Elizabeth M. George** is a detector engineer in the Detector Systems Group at European Southern Observatory. She develops instrumentation and detectors for astronomical telescopes from the visible to millimeter wavelengths. Previously, she completed her postdoctoral work on the ERIS project at the Max Planck Institute for Extraterrestrial Physics and her doctoral work in physics at UC Berkeley on the South Pole Telescope experiment.

**Simon Tulloch** is a detector engineer at the European Southern Observatory. He received his BS degree in Physics from the University of Kent and his PhD degree in Astrophysics from the University of Sheffield in 2010. He was previously head of the detector group at the Isaac Newton Group of telescopes in La Palma, Spain. His research interests include low-noise signal processing and photon counting detectors.

Biographies of the other authors are not available.